\documentclass[aps,twocolumn,showpacs,superscriptaddress]{revtex4-1}
\usepackage{amsfonts}
\usepackage{amsmath}
\usepackage{amssymb}
\usepackage{lipsum}
\usepackage{graphicx}
\usepackage{bm}
\usepackage{natbib}
\usepackage{float}
\usepackage{multirow}
\usepackage{ulem}
\usepackage{hyperref}
\usepackage{gensymb}
\hypersetup{colorlinks=true,
            linkcolor=blue,
            citecolor=blue,
            urlcolor=orange}
        
\usepackage[dvipsnames]{xcolor}

\usepackage{comment}
\def\comAK#1{{#1}}

\begin{document}
\preprint{APS/123-QED}
\title{Magnetic properties of monolayer, multilayer, and bulk CrTe$_2$}
\author{A. A. Katanin}
\affiliation{Center for Photonics and 2D Materials, Moscow Institute of Physics and Technology, Institutsky lane 9, Dolgoprudny, 141700, Moscow region, Russia}
\affiliation{M. N. Mikheev Institute of Metal Physics of the Ural Branch of the Russian Academy of Sciences, S. Kovalevskaya Street 18, 620990 Yekaterinburg, Russia}
\author{E. M. Agapov}
\affiliation{Ludwig-Maximilians-Universität München, Geschwister-Scholl-Platz 1, 80539, Munich, Germany}
\affiliation{Technical University of Munich, School of Natural Sciences, Arcisstraße 21, 80333, Munich, Germany}
\date{\today}
\begin{abstract}
We investigate magnetic properties of CrTe$_2$ within the density functional theory (DFT) approach in ferromagnetic phase and the combination of the DFT and the dynamical mean field theory (DFT+DMFT) approach in paramagnetic phase. We show that a few layer CrTe$_2$ possesses well formed local magnetic moments.
In the monolayer CrTe$_2$ we find the most preferable antiferromagnetic exchange with the 120$\degree$ antiferromagnetic structure. In the bilayer and trilayer systems, electronic correlations in DFT+DMFT approach yield ferromagnetic exchange interaction within each layer, but the interaction between the layers is antiferromagnetic, such that alternation of the direction of magnetization of the layers is expected. In bulk CrTe$_2$ we find the tendency to ferromagnetic order at low temperature, but with increase of temperature antiferromagnetic correlations between the layers dominate. Determination of the critical number of layers at which the interlayer antiferromagnetic order changes to the ferromagnetic likely requires consideration of the non-local Coulomb interactions. 
\end{abstract}
 \maketitle

\section{Introduction}
{
Atomically thin two-dimensional (2D) van der Waals (vdW) materials have attracted significant attention over the past decade due to their exceptional properties \cite{Rev1,Rev2,Rev3,Rev4,Rev5,Rev6}. In particular, they can posses large magnetic moments, substantial Curie temperature, comparable to room temperature, and strong anisotropy. With these properties, 2D magnetic materials are promising for using in the fields of spintronics \cite{Rev2} and memory storage devices \cite{Rev5}. } 
{Currently, many 2D ferromagnetic (FM) materials have been reported; however, a key challenge remains: achieving ferromagnetism at room temperature. According to the Mermin-Wagner theorem, a long-range magnetic order cannot exist in two dimensions in isotropic systems at finite temperatures due to thermal fluctuations. However, it is possible to circumvent this limitation by magnetic anisotropy.

Recently, chromium ditelluride CrTe$_2$ has been suggested to exhibit ferromagnetic order even up
to room temperature, making it an attractive candidate for further research in this field. 
Bulk CrTe$_2$ was shown \cite{freitas_ferromagnetism_2015} to posses FM order with $T_C = 330$~K 
and in-plane magnetic anisotropy.
Recent years several groups obtained thin CrTe$_2$ flakes via mechanical exfoliation \cite{purbawati_-plane_2020, sun_room_2020, fabre_characterization_2021}, chemical vapor deposition 
\cite{meng_anomalous_2021} and molecular beam epitaxy 
\cite{sun_ferromagnetism_2021, zhang_room-temperature_2021, liu_wafer-scale_2023, zheng_two-dimensional_2023}, showing stable FM order with $T_C \sim 200$~K. Although thick films were shown to have an in-plane magnetic anisotropy \cite{purbawati_-plane_2020, sun_room_2020, fabre_characterization_2021}, the films thinner 10 monolayers show out-of-plane magnetic anisotropy \cite{sun_ferromagnetism_2021, zhang_room-temperature_2021,meng_anomalous_2021}.
The authors of Refs. \cite{zhang_room-temperature_2021,meng_anomalous_2021,
xian_spin_2022, liu_wafer-scale_2023, 
wang_strain-_2024} succeeded fabricating monolayer devices. At the same time, the results of the magnetic state of monolayers remain controversial. Experimentally both, FM \cite{zhang_room-temperature_2021, meng_anomalous_2021, liu_wafer-scale_2023} and zigzag antiferromagnetic (AFM-zz) \cite{xian_spin_2022} states were reported. 
}

Theoretically, the density functional theory (DFT) approach was used to study various magnetic orders and exchange interactions in CrTe$_2$. In particular, FM \cite{lv_strain-controlled_2015, li_tunable_2021, liu_structural_2022, yang_tunable_2021, yao_control_2023}, AFM \cite{gao_thickness_2021,AFM2,abuawwad_noncollinear_2022}, charge density wave \cite{otero_fumega_controlled_2020} and incommensurate \cite{abuawwad_noncollinear_2022} states were considered as possible candidates in monolayer CrTe$_2$. \comAK{It was suggested in Refs.  \cite{wu_-plane_2022,zhu_insight_2023,gao_thickness_2021,lv_strain-controlled_2015,AFM2} that relatively small strain can cause the transition from AFM-zz to FM, which may explain different kinds of magnetic order observed in experiment. Furthermore, it was shown} in Ref. \cite{gao_thickness_2021} that the 
intralayer coupling of 2 to 4 layer CrTe$_2$ is still antiferromagnetic in DFT, similar to that of monolayer CrTe$_2$, and changes to ferromagnetic for a larger number of layers. At the same time, the interlayer exchange interaction was also found to change from AFM to FM between 4 and 5 layers. The change of the magnetic orders was related by the authors of Ref. \cite{gao_thickness_2021} to the change of the lattice constants from monolayer to bulk. \comAK{However, later study of Ref. \cite{wu_-plane_2022} has shown that at least in the bilayer system the in-plane antiferromagnetic order is stable in DFT for a wide range of lattice constants}.

\vspace{-0.08cm}
At the same time, since Cr is an open almost half-filled $d$ shell metal, it possesses strong correlations, which may affect preferable magnetic state.  Zhu et. al. \cite{zhu_insight_2023} showed that the effect of Hubbard interaction can be essential for the choice of magnetic phase in CrTe$_2$. Also, the authors of Ref. \cite{gao_thickness_2021} emphasized that the on-site Coulomb repulsion changes the interlayer coupling to AFM even in the bulk; important role of the Coulomb repulsion for bilayer systems was also discussed in Ref. \cite{Interlayer}. Previous theoretical works \cite{yao_control_2023, wu_-plane_2022, liu_structural_2022, 
li_tunable_2021,Interlayer} used DFT$+U$ approach to account for correlation effects. However, for metallic systems this approach may be insufficient. To treat the effect of correlations, we consider in the present study the DFT+DMFT approach. We use recently proposed method of evaluation of exchange interactions in the paramagnetic phase \cite{MyJ} to obtain an unbiased estimate of exchange interactions, which account for correlation effects and also consider their temperature dependence. We compare the results obtained with those from the DFT approach.

The plan of the paper is the following. In Section II we describe methods used for the magnetic properties of CrTe$_2$ investigation. In Section III we present results for temperature and momentum dependence of magnetic susceptibilities, exchange interactions for monolayer systems, followed by a study of bilayer and trilayer systems. In the end of the Section we consider bulk system to understand the limit of large number of layers.  Finally, in Sect. IV we present Conclusions.  
    
\section{Methods and parameters}

To obtain band structures, we use the DFT approach implemented in the Quantum Espresso \cite{QE} package with ultrasoft pseudopotentials from the SSSP PBEsol Precision library \cite{PPP},  supplemented by the maximally localized symmetry-adapted \cite{WanSym} Wannier projection on $3d$ Cr states and $5p$ Te states performed within the Wannier90 package \cite{Wannier90}. 

For bulk CrTe$_2$ we choose the experimental lattice parameters \cite{freitas_ferromagnetism_2015} $a_{\rm bulk}=3.7887$~\AA, $c_{\rm bulk}=6.0955$~\AA. 
The momentum grid $18\times18\times18$ was chosen.
For the monolayer CrTe$_2$ we choose the lattice parameter of Ref. \cite{abuawwad_noncollinear_2022} $a=3.71$~\AA~and consider interlayer spacing $c=20$~\AA~to avoid overlap of the orbitals between the layers; the momentum grid $20\times20\times1$ was chosen. The vertical position of Te atoms was relaxed and constituted the distance $1.55$~\AA~for the bulk and $1.62$~\AA~for the monolayer system from the Cr plane. 
For the bilayer (trilayer) CrTe$_2$ we choose the lattice parameter of Ref. \cite{liu_structural_2022} $a=3.76$~\AA~ ($a=a_{\rm bulk}$) and consider the unit cell of vertical size $c=40$~\AA~(60~\AA). The interlayer spacing was chosen equal to $c_{\rm bulk}$, Ref. \cite{sun_room_2020}. The relaxed vertical position of Te atoms constituted the distance $1.61$~\AA~ ($1.56$~\AA) from the Cr plane for the outer (inner) atoms for the bilayer system and $1.58$~\AA~ ($1.55$~\AA) for the trilayer system. The momentum grid was chosen $24\times24\times1$.

DMFT calculations of self-energies, non-uniform susceptibilities, and exchange interactions were performed within the continuous-time Quantum Monte Carlo (CT-QMC) method of the solution of impurity problem \cite{CT-QMC}, realized in the iQIST software \cite{iQIST}.
In DMFT calculation we consider the basis, which diagonalises crystal field, and use the density-density interaction matrix, parameterized by Slater parameters $F^0$, $F^2$, and $F^4$, expressed through Hubbard $U$ and Hund $J_H$ interaction parameters according to 
${F^0= U}$ and ${(F^2+F^4)/14 = J_{\rm H}}$, $F^2/F^4\simeq 0.63$ (see Ref.~\onlinecite{u_and_j}). The Coulomb interaction parameters were chosen $U=2.8$~eV, $J_H=0.9$~eV. We use a double-counting correction in the around mean-field form~\cite{AMF}.

To determine exchange interactions, we consider the effective Heisenberg model with the Hamiltonian $H=-(1/2)\sum_{{\bf q},rr'} J^{rr'}_{\bf q} {\mathbf S}^r_{\mathbf q} {\mathbf S}^{r'}_{-{\mathbf q}}$, {$\mathbf S^r_{\mathbf q}$ is the Fourier transform of static operators ${\mathbf S}_{ir}$} at the site $r$ of cell $i$,
where the orbital-summed static spin operators ${\mathbf S}_{ir}=\sum_m {\mathbf S}_{irm}$ 
The exchange interactions in DFT were estimated by magnetic force theorem (MFT)  approach \cite{korotin_calculation_2015-1,liechtenstein_local_1987, git} using electron Green's functions, calculated in Wannier basis 
$G_{rr', \sigma}^{m m^{\prime}}(\mathbf{k},i \nu_n)=\left[(i \nu_n + \mu)I-H^{rm, r'm^{\prime}}_{\mathbf{k},\sigma}\right]_{rm,r'm'}^{-1}$, where $H^{rm, r'm^{\prime}}_{\mathbf{k},\sigma}$ is the Wannier Hamiltonian, $I$ is the identity matrix, and the inversion is performed in the site- and orbital space, 
\begin{align}
J^{rr'}_{\mathbf{q}}&=  -\frac{2T}{{\mathfrak m}_r{\mathfrak m}_{r'}}  \sum_{{\mathbf k},i \nu_n}\operatorname{Tr}
    \left[\Delta_{ r} \tilde{G}_{ rr',\downarrow}({\mathbf k}+ {\mathbf q},i \nu_n)\right. \notag \\
    &\times \left.\Delta_{r'} \tilde{G}_{r'r,\uparrow}({\mathbf k},i \nu_n) \right] \label{JqDFT}
\end{align}
where $\Delta_{r}^{m m^{\prime}} = \sum_{\mathbf k}(H_{\mathbf{k},\uparrow }^{rm, rm^{\prime}}-H_{\mathbf{k},\downarrow}^{rm, rm^{\prime}})$ is spin splitting 
and $\tilde{G}^{mm'}_{ rr',\sigma}({\bf k},i \nu_n)={G}^{mm'}_{ rr',\sigma}({\bf k},i \nu_n)-
\sum_{{\mathbf k}'} G_{rr, \sigma }^{m m^{\prime}}(\mathbf{k}',i \nu_n)$, the trace is taken over orbital indexes, and $\mathfrak{m}_r$ is the magnetic moment at the $r$-th atom (in units of Bohr magneton $\mu_B$). 
For calculations we have used 200 Matsubara fermionic frequencies.


The exchange interactions in paramagnetic phase in DMFT were calculated  from the orbital-summed susceptibilities as \cite{MyJ,MyCo,Fe2C,CrO2}  $\hat{J}_{\mathbf{q}}={\hat{\chi}}_{\mathrm{loc}}^{-1}-{\hat{\chi}}_{\mathbf{q}}^{-1}$, where hats stand for matrices $n\times n$ in the atom number in the unit cell, $\chi^{rr'}_{\mathbf{q}}=-\langle\langle S_{\mathbf{q}}^{r,z} \mid S_{-\mathbf{q}}^{r',z}\rangle\rangle_{\omega=0}$  is the matrix of the non-local static longitudinal susceptibilities, containing local vertex corrections, and obtained from the respective Bethe-Salpeter equation \cite{MyEDMFT,OurRev} with account of 60-80 fermionic Matsubara frequencies, including also corrections to finite frequency box \cite{MyEDMFT} (at low temperatures we also use the method of Ref. \cite{Loon}),
$\chi^{rr'}_{\mathrm{loc}}=-\left\langle\left\langle S_i^{r,z} \mid S_i^{r,z}\right\rangle\right\rangle_{\omega=0}\delta_{rr'}$ is the diagonal matrix of local spin susceptibilities. 
The exchange interaction in real space is obtained by performing the Fourier transform of $J_{\mathbf q}$.

\section{Results}

\begin{figure}[t]
\centering
\includegraphics[width=0.62\linewidth,angle=-90]{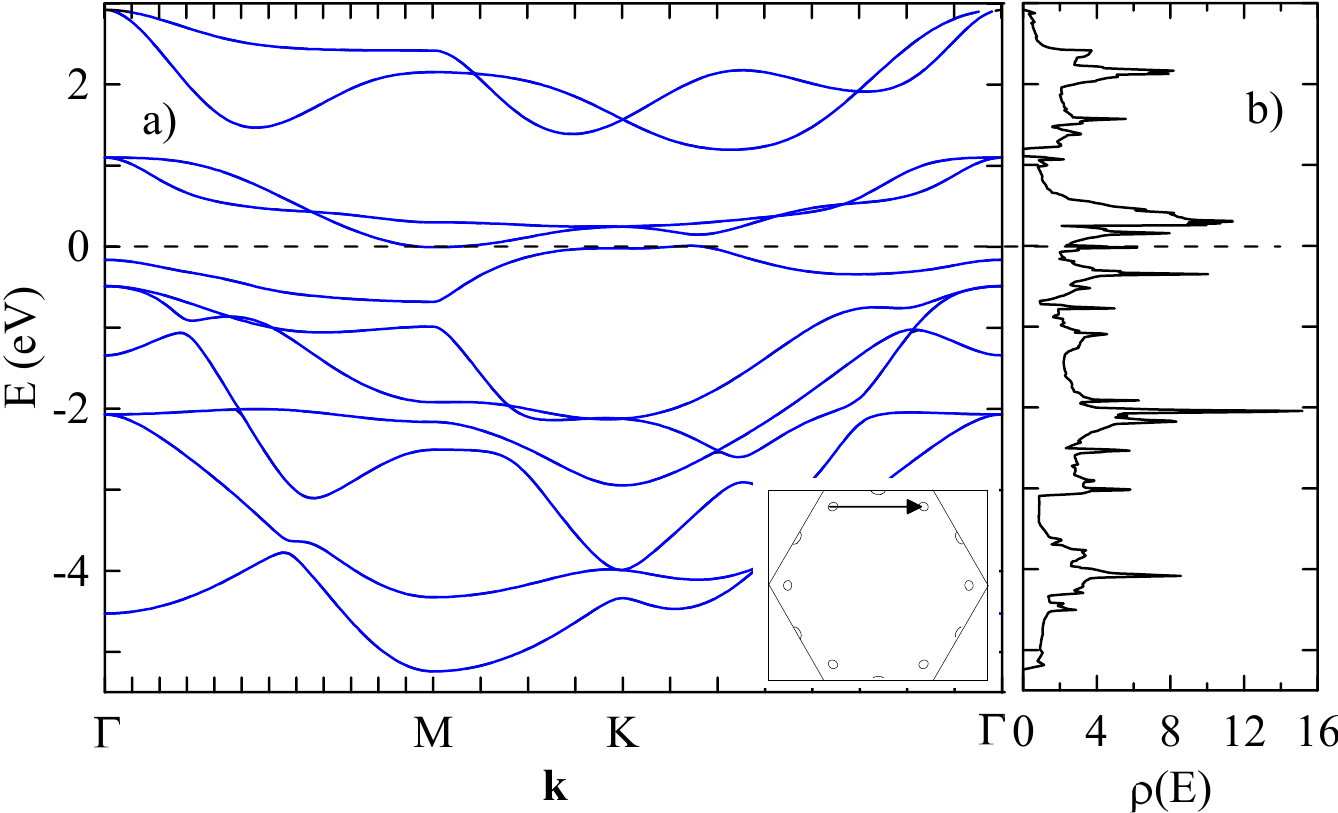}
\caption{(Color online) Band structure (a) and DFT density of states (b) of the monolayer CrTe$_2$ in paramagnetic phase. The inset in (a) shows the DFT Fermi surface; the arrow shows the nesting vector which is close to ${\mathbf Q}_K=\overrightarrow{\Gamma K}$.}
\label{Fig:bands_1_layer}
\end{figure}

The DFT band structure of the monolayer CrTe$_2$ in the paramagnetic state is shown in Fig. \ref{Fig:bands_1_layer}. In agreement with the earlier study \cite{sun_room_2020}, the density of states contains a weak peak at the Fermi level, which originates from the flat parts of the dispersion. The corresponding Fermi surface contains only small pockets near $K$, $K'$, and $M$ points of the Brillouin zone.

\begin{figure}[b]
\centering
\includegraphics[width=0.9\linewidth]{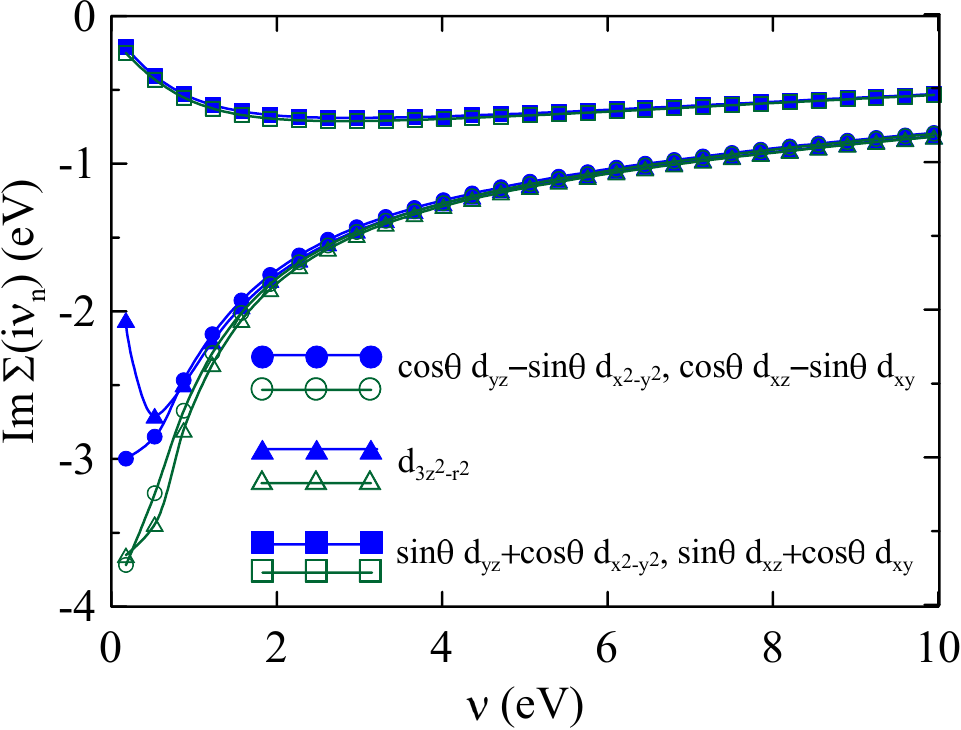}
\caption{Imaginary part of the self-energy of various orbital combinations of monolayer (blue solid symbols) and bilayer (green open symbols) CrTe$_2$ at the imaginary frequency axis in DFT+DMFT approach at $\beta=18$~eV$^{-1}$ \comAK{($T=645$~K)}.}
\label{Fig:Sigma}
\end{figure}

The filling of the chromium d orbitals, obtained in DFT+DMFT calculations for the monolayer compound CrTe$_2$, is $n_{\rm Cr}=4.83$, and therefore sufficiently close to half filling, which provides rather strong correlation effects. In Fig. \ref{Fig:Sigma} we show imaginary parts of the electronic self-energy for combinations of orbitals, which diagonalize crystal field (the rotation angle $\theta\simeq \pi/3$). One can see that the for some of the states (those which are close to the Fermi level) the damping of electronic excitations is particularly large, and the self-energy for some of the combinations has a non-quasiparticle form with $\partial {\rm Im}\Sigma/\partial \nu>0$.

\begin{figure}[t]
\centering
\includegraphics[width=0.85\linewidth]{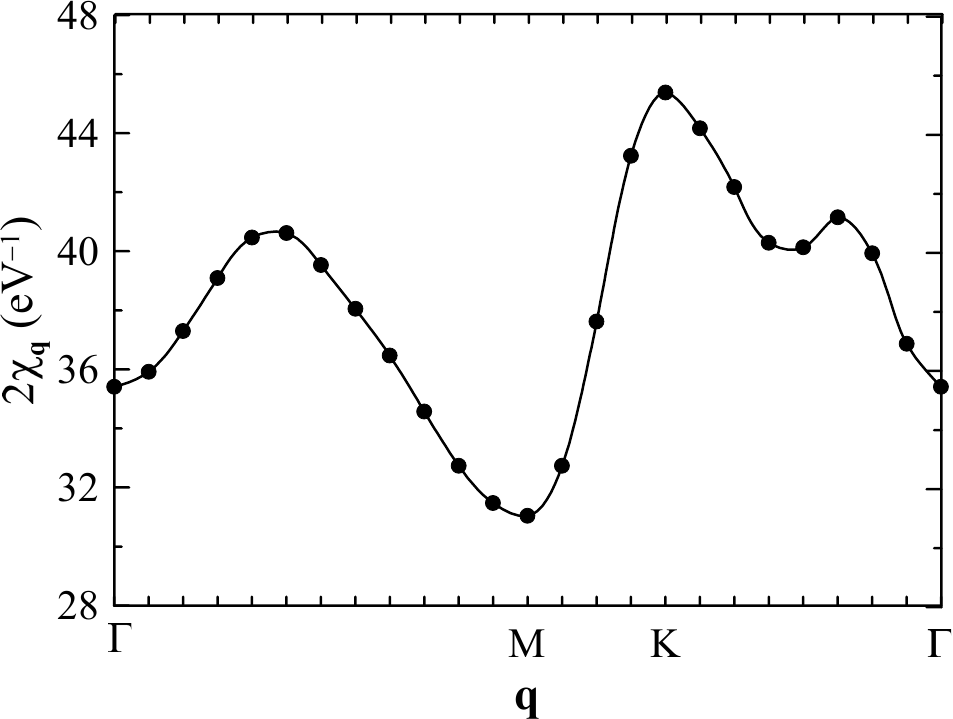}
\caption{Momentum dependence of the orbital-summed susceptibility of the monolayer CrTe$_2$ in DFT+DMFT approach at $\beta=7$~eV$^{-1}$ \comAK{($T=1660$~K)}. }
\label{Fig:chiq_1_layer}
\end{figure}

The magnetic susceptibility obtained at $\beta=7$~eV$^{-1}$ \comAK{($T=1660$~K)} is rather large (see Fig. \ref{Fig:chiq_1_layer}) and has a maximum at the wave vector $\mathbf{Q}_K=(4\pi,0)/(3a)$, corresponding to the tendency to the $120\degree$ incommensurate magnetic order. Despite the strong incoherence of electronic excitations, this tendency is likely related to the nesting of the pockets of the Brillouin zone, as shown in the inset of Fig. \ref{Fig:bands_1_layer}a. The susceptibility does not show even a local maximum near the $\Gamma$ point, which implies an absence of pronounced tendency towards ferromagnetic order.

\begin{figure}[b]
\centering
\includegraphics[width=0.47\linewidth,angle=-90]{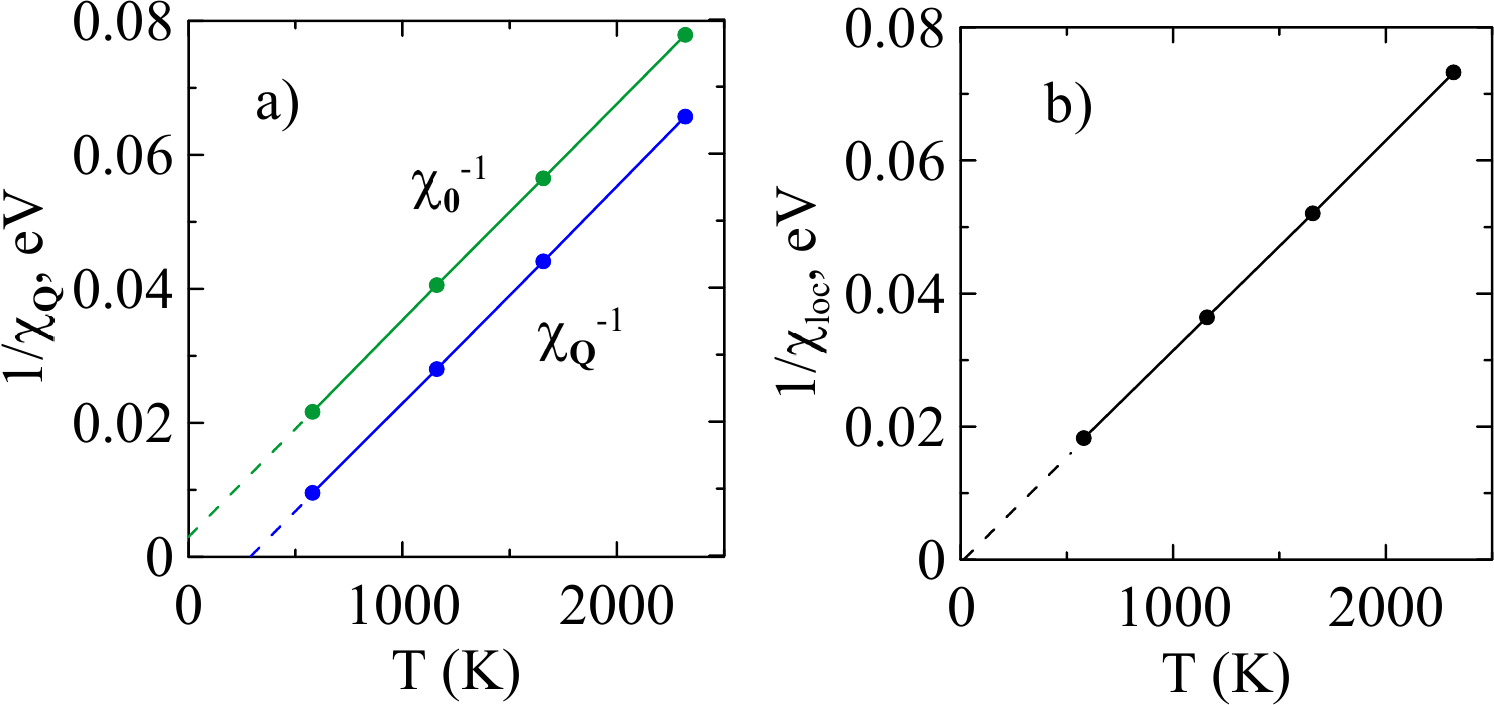}
\caption{(Color online) Inverse uniform and staggered, ${\mathbf Q}={\mathbf Q}_K$ (a) and local (b) spin susceptibility of the monolayer CrTe$_2$ in DFT+DMFT approach. Dashed lines show extrapolation to the low-temperature region. }
\label{Fig:chi_1_layer}
\end{figure}

The temperature dependence of the inverse uniform and staggered magnetic susceptibilities is shown in Fig. \ref{Fig:chi_1_layer}. The corresponding DMFT Neel temperature, obtained by extrapolating the inverse staggered susceptibility, $T^{\rm DMFT, 1{\text -}layer}_N\simeq 300 K$. At the same time, the inverse uniform susceptibility is extrapolated to the small negative Weiss temperature, showing absence of the ferromagnetic order, in agreement with the conclusion from the momentum dependence of the nonuniform susceptibility. The inverse local susceptibility (see Fig. \ref{Fig:chi_1_layer}b) is almost linear in temperature, showing well-formed local magnetic moments with negligibly small Kondo temperature, estimated as an offset of linear extrapolation from zero. The local magnetic moment, obtained from the slope of the local (staggered) susceptibility $\mu_{\rm loc}^2=32.4 \mu_B^2$ ($\mu^2=31.7 \mu_B^2$), corresponding to the local effective spin $S_{\rm eff}\simeq 2.4$ defined by $\mu^2=(g\mu_B)^2 S_{\rm eff}(S_{\rm eff}+1)$, $g=2$ is the spin $g$-factor.

\begin{figure}[t]
\centering
\includegraphics[width=0.9\linewidth]{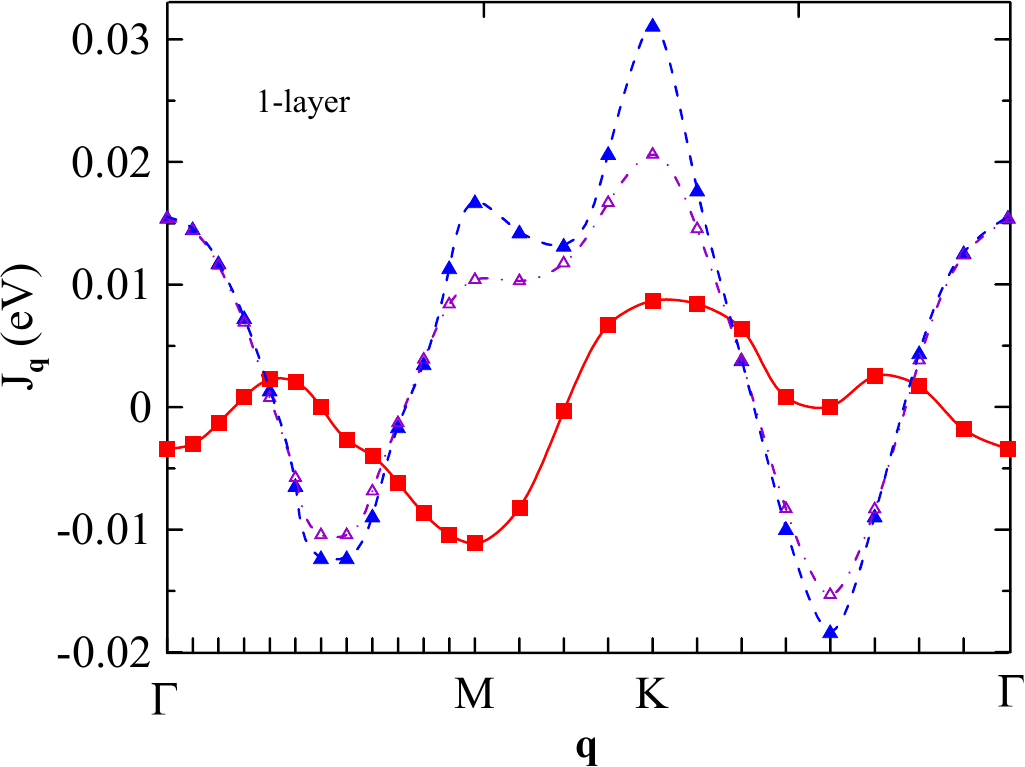}
\caption{(Color online) Exchange interaction in monolayer CrTe$_2$ at $\beta=20$~eV$^{-1}$ \comAK{($T=580$~K)}. Red solid line (squares) corresponds to DFT+DMFT approach in paramagnetic phase, blue dashed line (filled triangles) is the result of the DFT approach in FM phase. For comparison, the result of DFT approach at $\beta=10$~eV$^{-1}$ \comAK{($T=1160$~K)} is shown by dot-dashed violet line with open triangles}
\label{Jq_1_layer}
\end{figure}

Magnetic exchange interactions of the monolayer CrTe$_2$ at $\beta=20$~eV$^{-1}$ \comAK{($T=580$~K)} are shown in Fig. \ref{Jq_1_layer}. The maximum of exchange interaction in both, the DFT and DFT + DMFT approaches is also found at the wave vector ${\bf Q}_K$, which corresponds to the 120$\degree$ antiferromagnetic order.  
FM order is locally stable in DFT due to the local maximum at the $\Gamma$ point with $J_0 > 0$ and unstable in DMFT, but in both cases the $\Gamma$ point is not a global maximum, which corresponds to an instability of ferromagnetism in Heisenberg model.
  We find that the peak of exchange interaction $J_{\mathbf q}$ at $\mathbf{q}=\mathbf{Q}_K$ in DFT approach sharply increases with a decrease of temperature (see the results of $\beta=10$~eV$^{-1}$ in Fig. \ref{Jq_1_layer} for comparison), due to nesting of the Fermi surface pockets with the wave vector close to $\mathbf{Q}_{K}$ (see Appendix A). At the same time, the exchange interactions in the DFT+DMFT approach are quite weakly temperature-dependent because of the damping of electronic excitations.  We also note that in the ferromagnetic ordered phase of DFT we obtain the magnetic moment of chromium sites $\mathfrak{m}=2.86\mu_B$, which corresponds to the effective spin $S_{\rm ord}=\mathfrak{m}/(g\mu_B)=1.48$. This reflects different spin states of chromium in the ferro- and paramagnetic phases: while the chromium occupation is close in two approaches ($n_{\rm Cr}=4.67$ in DFT), the effective spin difference $\Delta S=S_{\rm eff}-S_{\rm ord}\simeq 1$ corresponds to only partial magnetization of orbital states. In particular, we find the strongest magnetization contribution $0.8\mu_B$ / f.u. from the $d_{3z^2-r^2}$ state and the weakest contribution $0.4\mu_B$ / f.u. from each of the $d_{xz,yz}$ states. 

\begin{table}[b]
\begin{tabular}{lrrr}
Method                     & $\ \ J_1$\ \  & $J_2$\ \   & $J_3$ \ \ \\
\hline
DFT Energies, this work & $\comAK{-1.52}$   & $\comAK{1.02}$ &  $\comAK{-0.17}$    \\
DFT MFT, this work & $-1.51$   & $3.52$ &  $2.06$    \\
DMFT, this work               & \ \ $0.08$    & $1.16$  & $-1.34$ \\
DFT Energies \cite{li_tunable_2021}               & \ \ 5.20    & 0.94  &      \\
 all-electron KKR-GF \cite{abuawwad_noncollinear_2022}    & $-3.19$     & $2.37$   & $1.18$
\end{tabular}
\caption{Exchange interactions (in meV) in monolayer CrTe$_2$ for nearest ($J_1$), next nearest ($J_2$) and next to next nearest ($J_3$) neighbours. DFT MFT and DMFT exchange interactions are calculated at $\beta=20$~eV$^{-1}$ \comAK{($T=580$~K)}.} 
\label{Table:exchenge_monolayer}
\end{table}

\begin{figure}[t]
\centering
\includegraphics[width=0.62\linewidth,angle=-90]{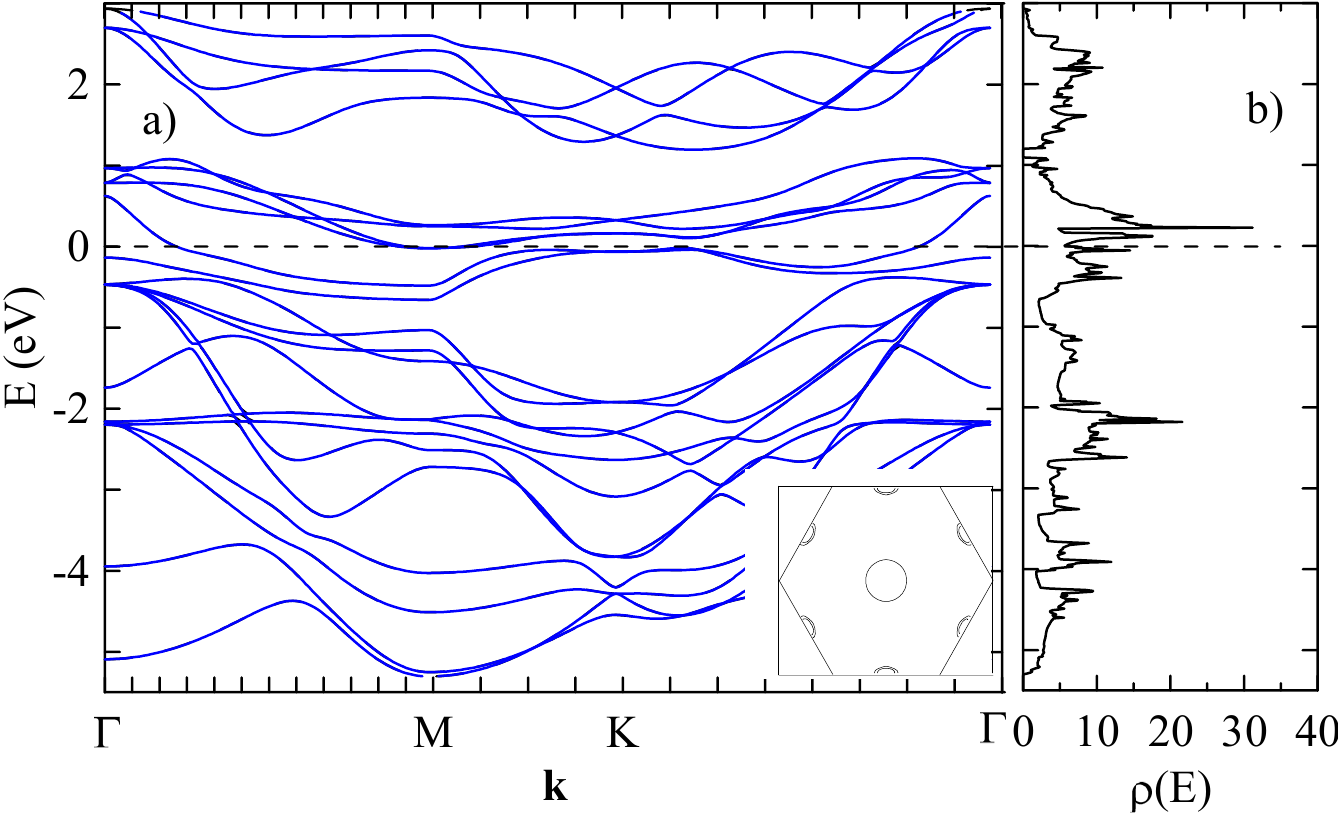}
\caption{(Color online) Band structure (a) and DFT density of states (b) of the bilayer CrTe$_2$ in paramagnetic phase. The inset in (a) shows the DFT Fermi surface.}
\label{Fig:bands_2_layer}
\end{figure}

We also used an alternative method 
to obtain exchange interactions, based on the comparison of energies of various spin configurations \cite{li_tunable_2021}, see Appendix B. The comparison of the obtained exchange interactions is presented in Table \ref{Table:exchenge_monolayer}. All DFT approaches, except Ref. \cite{li_tunable_2021} yield the same signs of nearest- and next-nearest neighbor exchange interactions, with different magnitude. In particular, the nearest neighbour exchange interaction is negative (except Ref. \cite{li_tunable_2021}) showing the tendency to antiferromagnetism. We find the sAFM-zz configuration in DFT approach 10 meV/f.u. lower than FM. At the same time, in DMFT approach we find weak ferromagnetic nearest neighbor exchange interaction, and the tendency to antiferromagnetism originates from the next to next nearest neighbors and further interactions. 

\begin{figure}[t]
\centering
\includegraphics[width=0.85\linewidth]{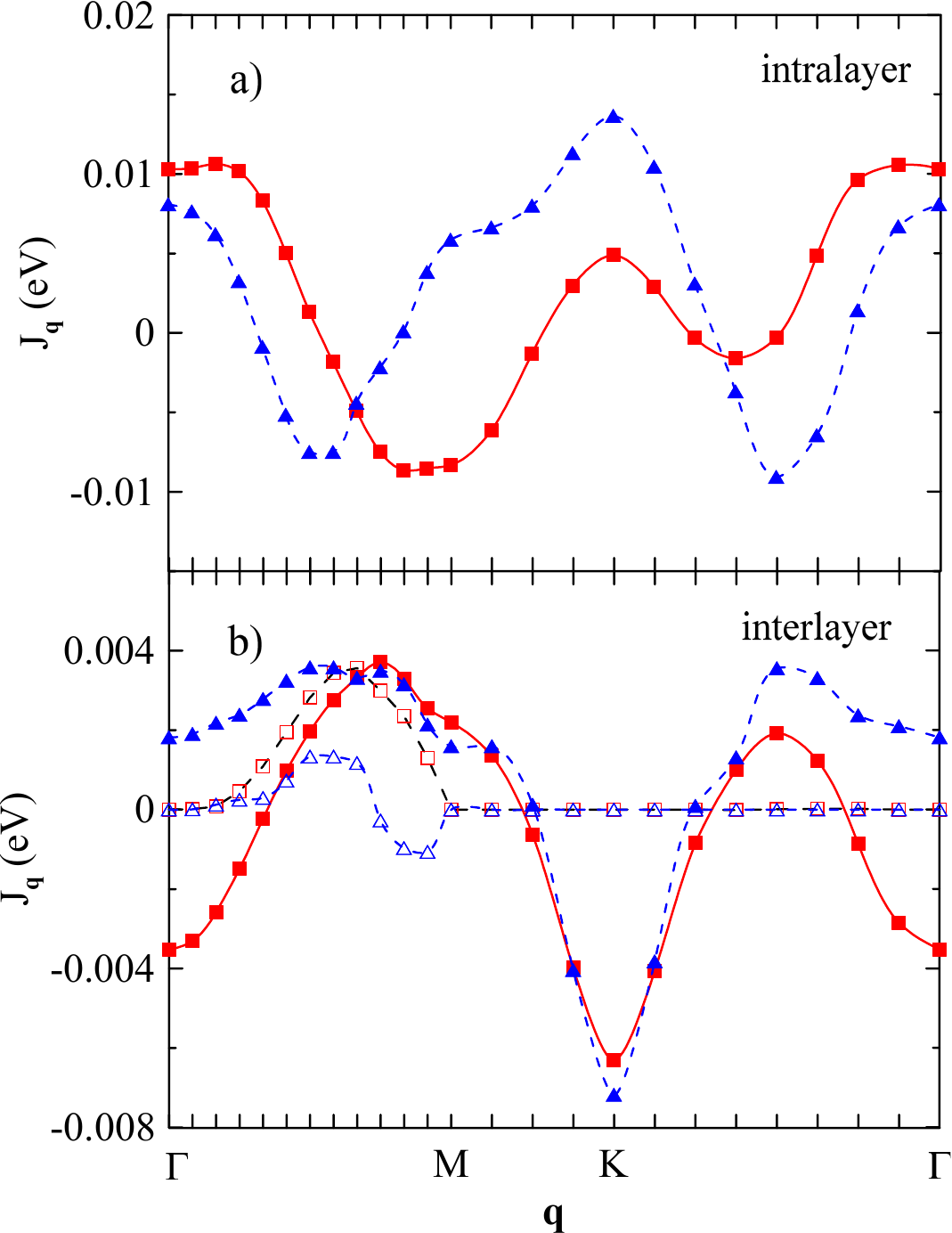}
\caption{(Color online) Exchange interaction in bilayer CrTe$_2$ at $\beta=18$~eV$^{-1}$ \comAK{($T=645$~K)} (a) between the atoms of the same layer $J^{11}_{\mathbf q}$ and (b) between the atoms of different layers $J^{12}_{\mathbf q}$. Red solid line (squares) corresponds to DFT+DMFT approach in paramagnetic phase, blue dashed line (triangles) is the result of the DFT approach in FM phase. Full (open) symbols in (b) correspond to the real (imaginary) part. }
\label{Fig:Jq_2_layer}
\end{figure}

\begin{figure}[t]
\centering
\includegraphics[width=0.95\linewidth]{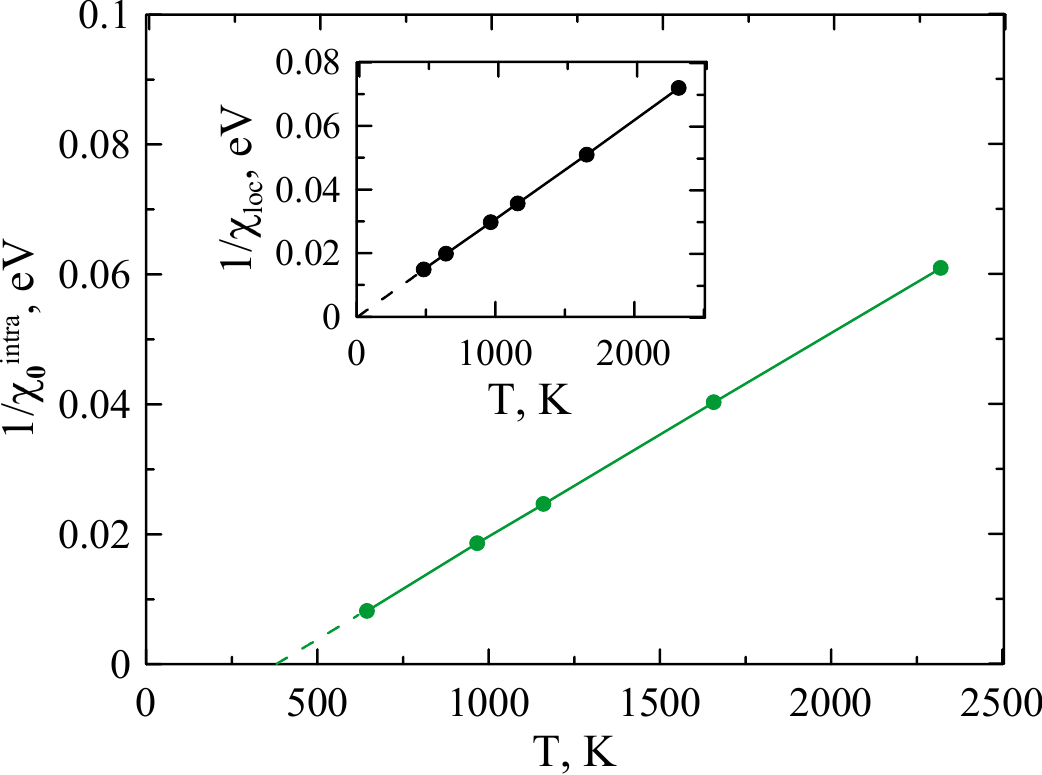}
\caption{(Color online) Inverse uniform intralayer susceptibility (main plot) and local susceptibility (inset) of the bilayer CrTe$_2$ in DFT+DMFT approach. Dashed lines show extrapolation to the low-temperature region. }
\label{Fig:chi_2_layer}
\end{figure}

To study the effect of interlayer coupling we consider bilayer and trilayer CrTe$_2$. The band structure of bilayer CrTe$_2$ in paramagnetic phase is shown in Fig. \ref{Fig:bands_2_layer}. One can see that due to bilayer splitting the central sheet of the Fermi surface appears at the $\Gamma$ point. The electronic self-energies are shown in Fig. \ref{Fig:Sigma}. One can see that the electronic damping for the bilayer system is even larger than for the monolayer one, and the non-quasiparticle form of the self-energy is also more pronounced.


\begin{figure}[b]
\centering
\includegraphics[width=0.61\linewidth,angle=-90]{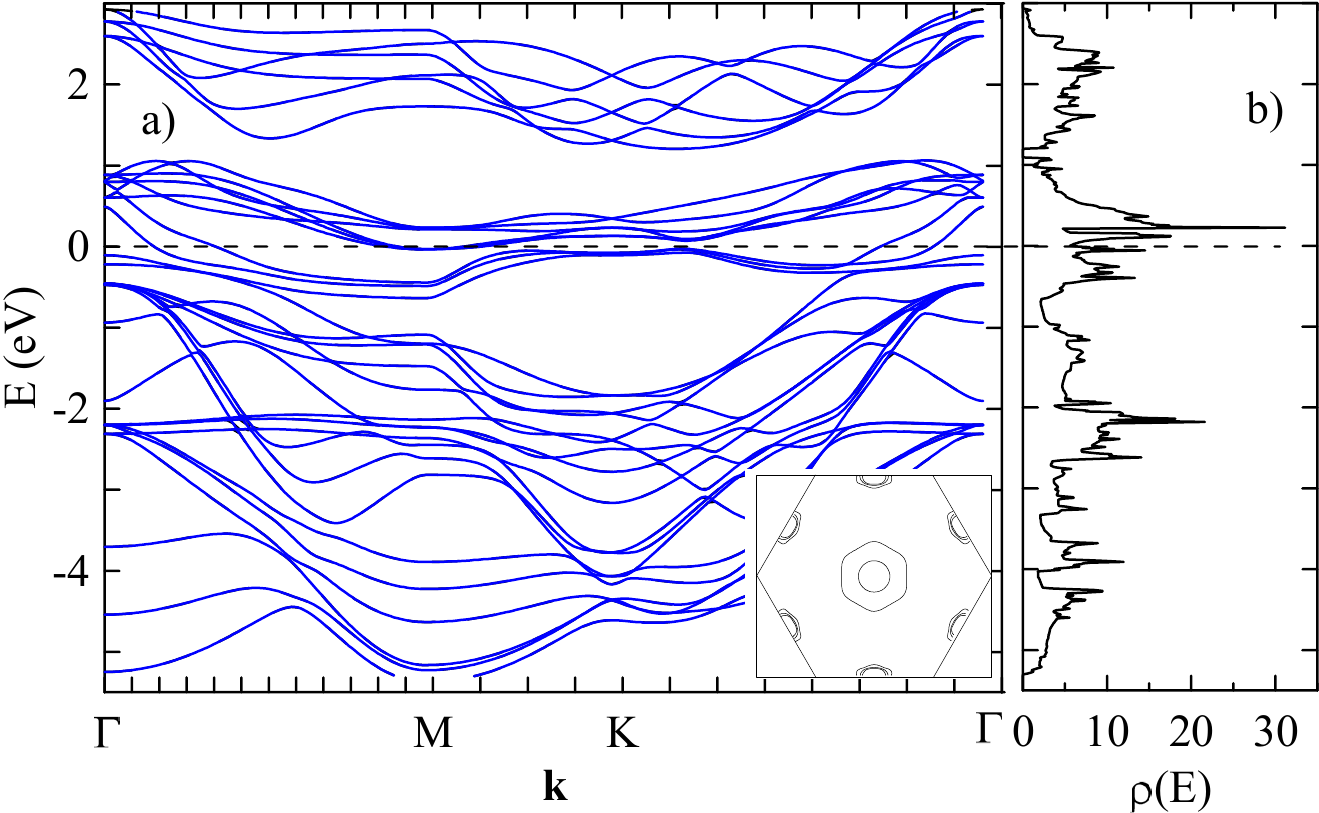}
\caption{(Color online) Band structure (a) and DFT density of states (b) of the trilayer CrTe$_2$ in paramagnetic phase. The inset in (a) shows the DFT Fermi surface.}
\label{Fig:bands_3_layer}
\end{figure}

We consider the interaction between the atoms of the same Cr plane (intralayer part), $J^{rr}_{\mathbf q}$ and the interplane interaction $J^{rr'}_{\mathbf q}$ with $r\neq r'$. 
From the momentum dependence of intralayer exchange interactions $J^{11}_{\mathbf q}=J^{22}_{\mathbf q}$ (see Fig. \ref{Fig:Jq_2_layer}a) one can see that an additional sheet of the Fermi surface strongly affects exchange interactions: the intralayer interaction becomes maximal near the $\Gamma$ point of the Brillouin zone, while the height of the local maximum at the $K$ point decreases in comparison to the monolayer system. 
The obtained maximum of the intralayer exchange interaction can be contrasted to the results of DFT approach of the present study (shown by dashed line in Fig. \ref{Fig:Jq_2_layer} a) and the earlier DFT study \cite{zhu_insight_2023}, where the antiferromagnetic ground state was obtained. 
In case of bilayer system, both the DFT and the DFT+DMFT results for exchange interactions are weakly temperature dependent because of the absence of nesting and shift of the peak of the density of states from the Fermi level. \comAK{We have also verified that the tendency to the in-plane ferromagnetic order in DFT+DMFT approach in paramagnetic phase is stable for $a>3.73$\AA}.

\begin{figure}[t]
\centering
\includegraphics[width=0.95\linewidth]{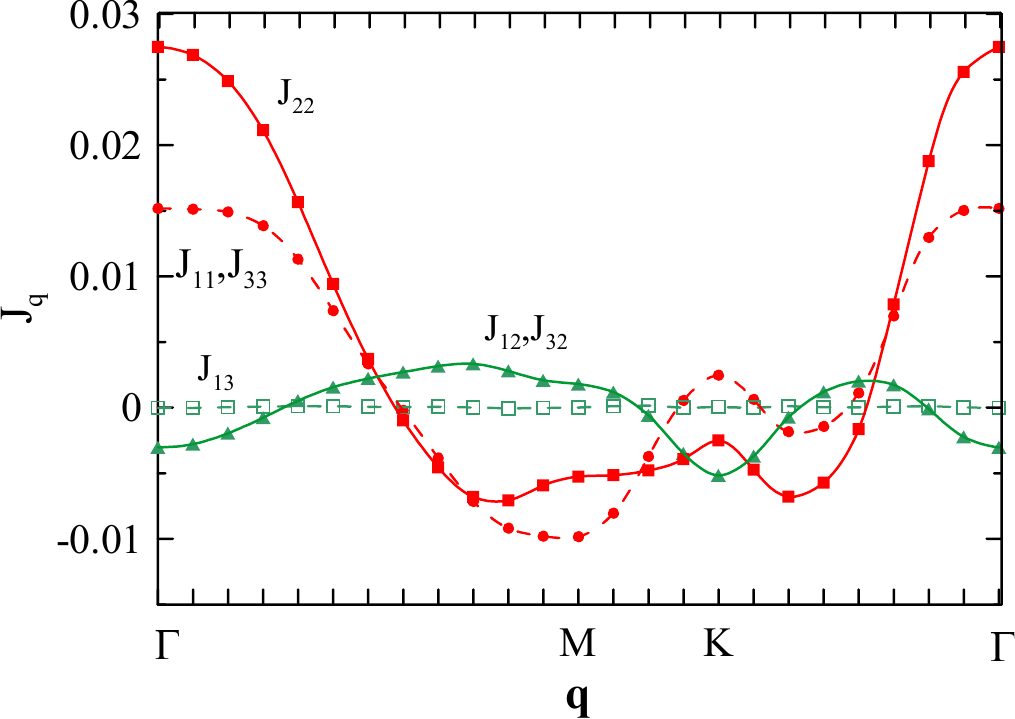}
\caption{(Color online) Real parts of the exchange interactions $J_{ij}({\mathbf q})\equiv J^{ij}_{\mathbf q}$ in trilayer system at $\beta=18$~eV$^{-1}$ {(645 K)} in DFT+DMFT approach. The pairs of indexes $ij$ at the exchange interactions correspond to the pairs of layers (1-bottom, 2-middle, 3-top layer)}
\label{Fig:Jq_3_layer}
\end{figure}

To \comAK{further} demonstrate the possibility of intralayer ferromagnetic order in the bilayer CrTe$_2$, we also show the temperature dependence of the inverse uniform intralayer susceptibility $(\chi^{11,22}_{{\mathbf q}=0})^{-1}$ in Fig. \ref{Fig:chi_2_layer}. One can see that in contrast to monolayer CrTe$_2$ the inverse intralayer susceptibility vanishes at the DMFT Curie temperature $T_{C}^{\rm DMFT,  {\text 2{\text -}layer}}\simeq 400K$. Similarly to the monolayer system, this DMFT Curie temperature does not determine the true Curie temperature of the system, but rather the scale of the onset of strong ferromagnetic correlations.   The temperature dependence of the local magnetic susceptibility (see the inset of Fig. \ref{Fig:chi_2_layer}) is similar to the monolayer case, and it is characterized by the magnetic moment $\mu^2_{\rm loc}=33.1\mu_B^2$, which is also close to that for the monolayer compound. Therefore the strong change of the non-local properties when passing from single- to bilayer CrTe$_2$ almost does not affect the local properties.

The interlayer exchange interaction $J^{12}_{\mathbf q}=(J^{21}_{\mathbf q})^*$ of the bilayer system is negative at the $\Gamma$-point (see Fig. \ref{Fig:Jq_2_layer}b), showing the tendency to opposite orientation of spins in the layers,  
similarly to the previous DFT+$U$ studies  \cite{gao_thickness_2021,Interlayer}, while the DFT approach yields positive $J^{12}_{\mathbf 0}$. Nevertheless, both approach yield maximum of the interlayer exchange interaction at the $\Gamma$-$M$ and $\Gamma$-K directions, which corresponds to the onset of incommensurate magnetic order between the layers.  We note that to the best of our knowledge, at the moment there are no experimental data on bilayer systems available.

Let us consider what changes in DFT+DMFT results when adding one more layer in trilayer CrTe$_2$. The Fermi surface possesses on additional sheet near the $\Gamma$ point (see Fig. \ref{Fig:bands_3_layer}). We find that the self-energies of trilayer system (not shown) are comparable to those of the bilayer one in Fig. \ref{Fig:Sigma}. The maximum of the intralayer exchange interaction (see Fig. \ref{Fig:Jq_3_layer}) shifts to the $\Gamma$ point (instead of being in the near vicinity of the $\Gamma$ point in bilayer system), which shows further stabilization of intralayer ferromagnetic order with increase of the number of layers. 
The exchange interaction between adjacent layers is negative (i.e. antiferromagnetic), similar to the bilayer compound, with the maximum of the interlayer exchange interactions at the incommensurate positions in the $\Gamma-M$ and $\Gamma-K$ directions. \comAK{We find that the tendency to incommensurate interlayer order is stable with further lowering the temperature.}

\begin{figure}[b]
\centering
\includegraphics[width=0.61\linewidth,angle=-90]{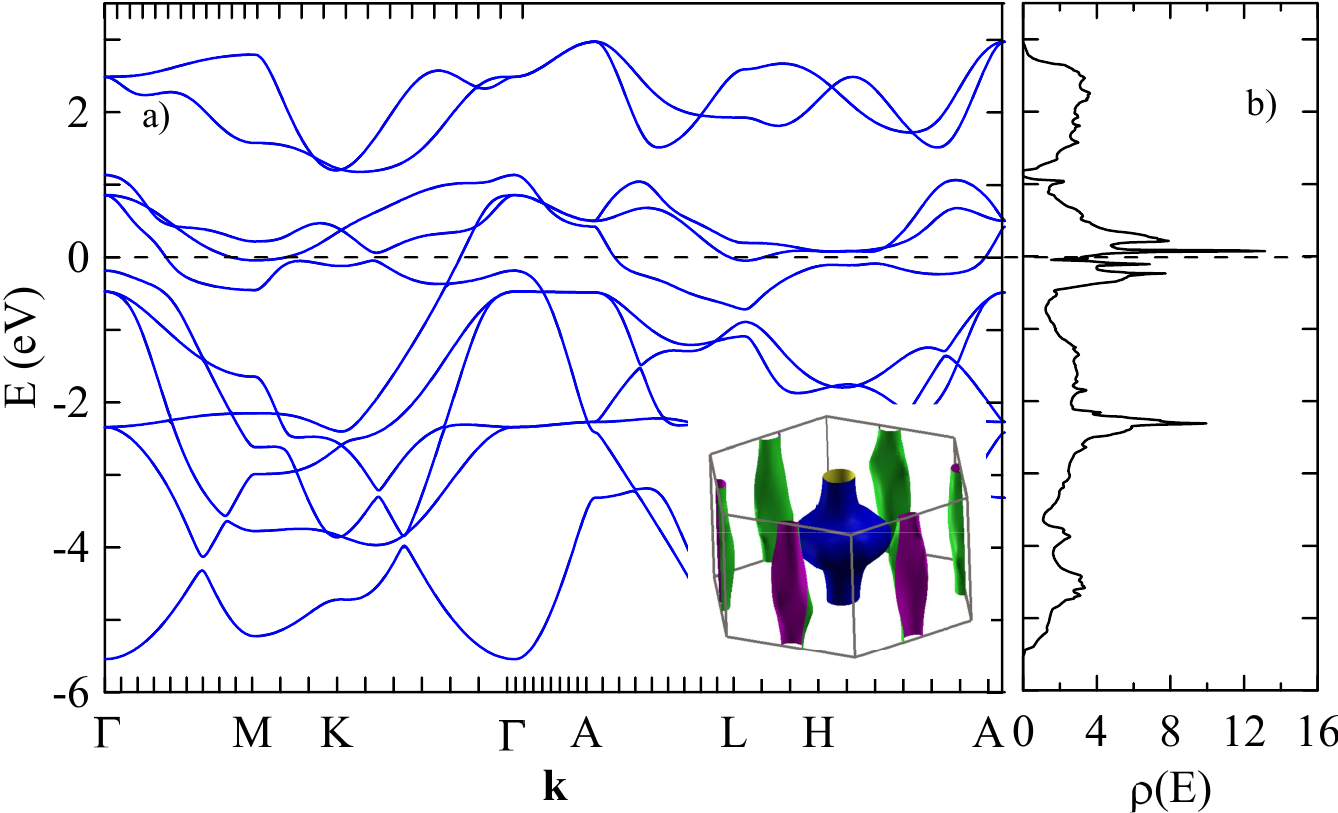}
\caption{(Color online) Band structure (a) and DFT density of states (b) of bulk CrTe$_2$ in paramagnetic phase. The inset in (a) shows the DFT Fermi surface.}
\label{Fig:bands_bulk}
\end{figure}

\begin{figure}[t]
\centering
\includegraphics[width=0.9\linewidth]{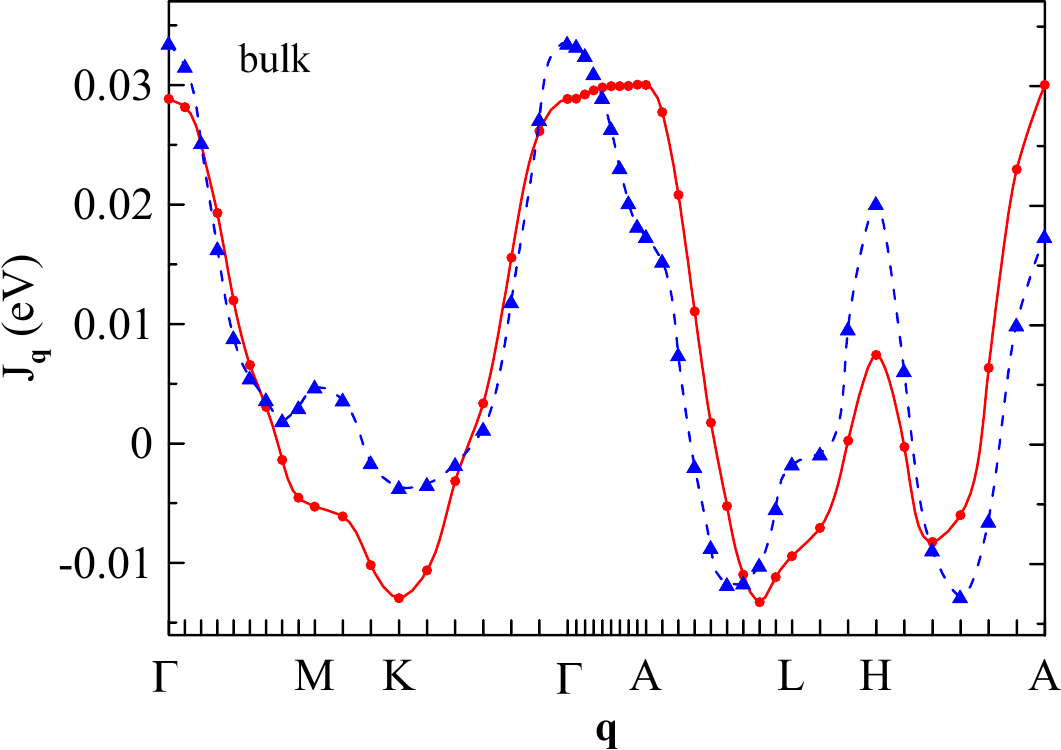}
\caption{(Color online) Exchange interaction in bulk CrTe$_2$ at $\beta=24$~eV$^{-1}$ \comAK{($T=480$~K)} in DFT+DMFT approach in paramagnetic phase \comAK{(red solid line with circles) and the DFT approach in the FM phase (blue dashed line with filled triangles).}}
\label{Fig:Jq_bulk}
\end{figure}
\begin{figure}[b]
\centering
\includegraphics[width=0.9\linewidth]{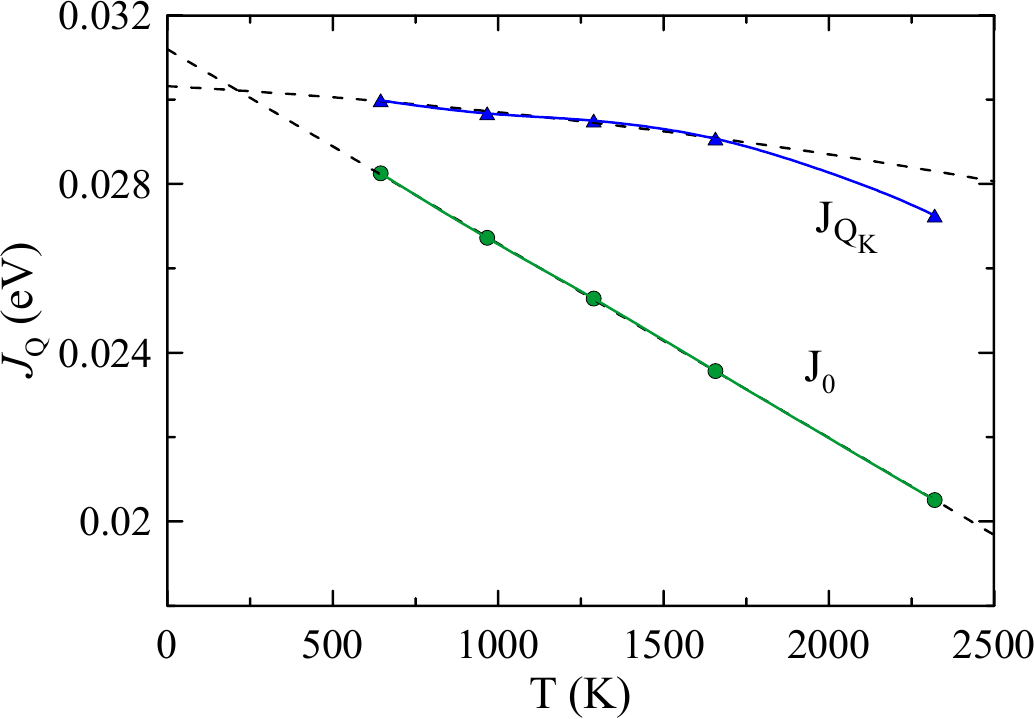}
\caption{(Color online) Temperature dependence of exchange interactions $J_0$ (circles) and $J_{{\mathbf Q}_K}$ (triangles) in bulk CrTe$_2$. Dashed lines show the result of extrapolation.}
\label{Fig:Jq_bulk1}
\end{figure}

To understand the effect of adding more layers in thick films, we consider bulk CrTe$_2$. The obtained band structure and Fermi surface (Fig. \ref{Fig:bands_bulk}) reproduce those of previous study \cite{freitas_ferromagnetism_2015}. The Fermi surface sheet closest to the $\Gamma$ point acquires the dispersion along $k_z$ axis in agreement with the results for the trilayer system. The exchange interactions at $\beta=24$~eV$^{-1}$ \comAK{($T=480$~K}, see Fig. \ref{Fig:Jq_bulk}) has local maximum at the $\Gamma$ point, but its value at the A point of the Brillouin zone is somewhat larger than at $\Gamma$, showing the tendency to alternating orientation between adjacent planes, similar to that obtained for 2- and 3-layer system. However, with a reduction of temperature (see Fig. \ref{Fig:Jq_bulk1}), we observe a clear tendency of changing the global maximum of exchange interaction to $\Gamma$ point, which happens just at $T_C\simeq 250$~K, comparable to the experimental Curie temperature. This implies formation of low temperature FM phase (in agreement with the experimental data), which changes upon heating to a state with long- or short-range antiferromagnetic order. 
\comAK{From the obtained exchange interactions, we find using RPA approach \cite{RPA_TC} (see also Refs. \cite{MyCo,CrO2}), that the Neel temperature of antiferromagnetic state  approximately coincides with the obtained $T_C$.} Therefore, we expect that the long-range antiferromagnetic order above $T_C$ is destroyed by fluctuations.  Consequently, we find an intriguing physical picture of competition of different types of magnetic correlations, 
which occurs just above the Curie temperature. 


In view of the obtained results for the bilayer and trilayer systems, with decrease of the number of layers we expect the interlayer antiferromagnetic and incommensurate orders to become progressively more favorable. Therefore, we expect the transition to interlayer incommensurate or antiferromagnetic states with a decrease of the number of layers. In view of the delicate competition of various states, additional interactions, not included in the considered model (such as $d$-$p$ and $p$-$p$ interactions) may be important for obtaining the critical number of layers for this transition, cf. Ref. \cite{Interlayer}.
\vspace{-0.15cm}
\section{Conclusions}
\vspace{-0.15cm}
In summary, we have investigated magnetic properties of monolayer, bilayer, trilayer, and bulk CrTe$_2$. In monolayer system in both the DFT and DFT + DMFT approaches, we find a preference for the incommensurate magnetic order with the wave vector ${\mathbf Q}_K$, which corresponds to 120\degree~ spin alignment. This order competes with the AFM-zz magnetic order, previously suggested for monolayer system.

In the bilayer and trilayer CrTe$_2$ in the DFT+DMFT approach, we find, in contrast to the DFT results, the tendency to the intralayer ferromagnetic order, which appears due to electronic correlations. At the same time, similarly to previous DFT$+U$ studies, we find a tendency to the interlayer AFM order. This tendency is preserved at not too low temperatures in bulk CrTe$_2$, and changes to FM order at low temperatures. 

The most important result of the present study is the possibility of stabilizing the intralayer FM order in CrTe$_2$ by correlations for the number of layers greater than one. Description of the transition from FM to AFM interlayer coupling with the decrease of the number of layers requires considering more sophisticated models, including non-local $d$-$p$ (as well as $p$-$p$)  interaction. The FM order in monolayer CrTe$_2$, observed experimentally \cite{zhang_room-temperature_2021, meng_anomalous_2021, liu_wafer-scale_2023}, requires further studies, and likely appears as an effect of substrate, cf. Ref. \cite{wang_strain-_2024}. 

For future studies, account of full SU(2) Coulomb interaction as well as the effect of the spin-orbit coupling would be desirable. Also, considering the effect of substrate, e.g., within the recently proposed approach of Ref. \cite{Substr} is of certain interest.

\section*{Acknowledgements} We are grateful to I. A. Goremykin for help with the Quantum Espresso and Wannier90 packages and discussions, and to A. K. Nukhov for discussions of the obtained results. The paramagnetic DFT+DMFT calculations and their analysis were supported by the Russian Science Foundation (Project No. 24-12-00186). The development of a program of finite frequency box corrections to the Bethe-Salpeter equation is supported within the theme “Quant” 122021000038-7 of Ministry of Science and Higher Education of the Russian Federation. 


\appendix

\section{Monolayer band structure in the ferromagnetic phase}
\begin{figure}[t]
\centering
\includegraphics[width=0.61\linewidth,angle=-90]{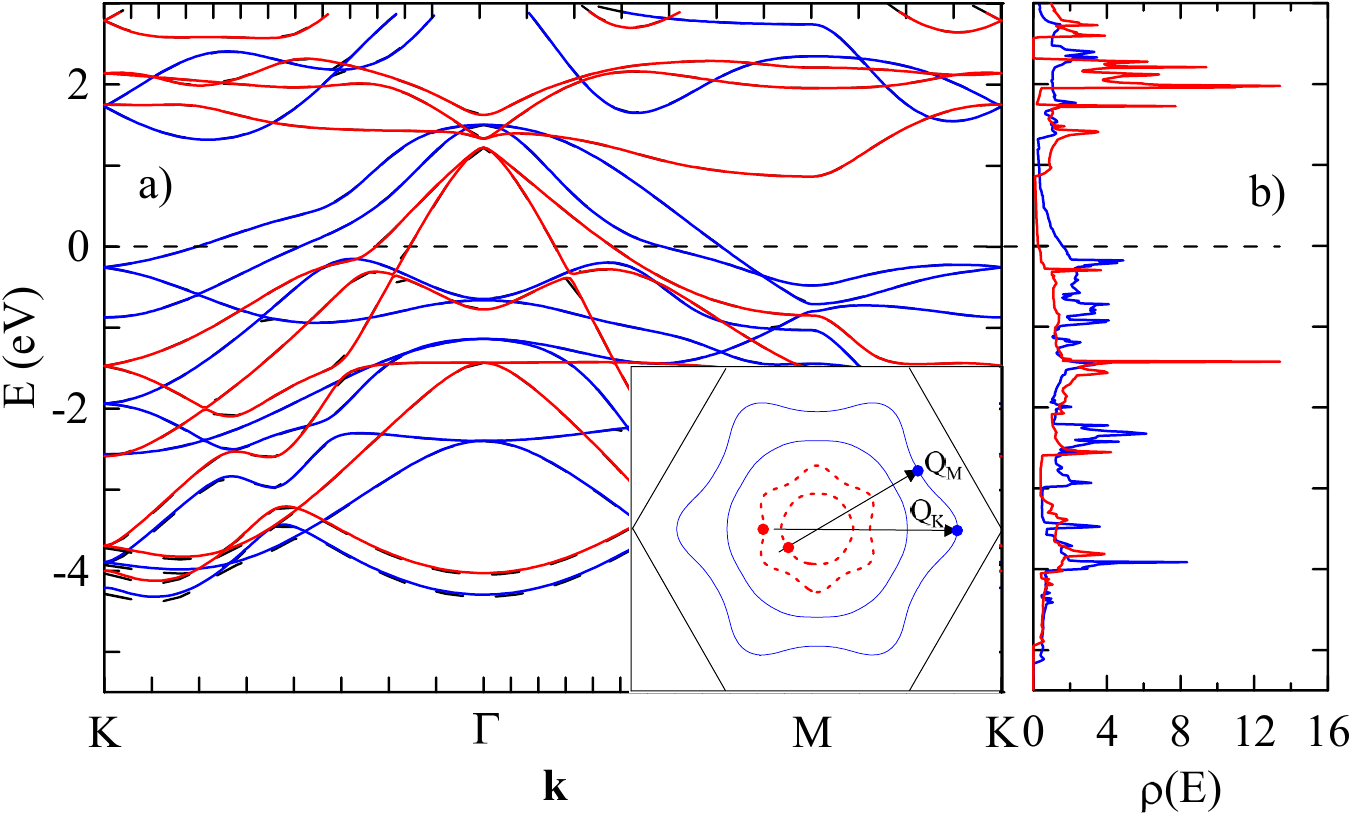}
\caption{(Color online) Band structure (a) and DFT density of states (b) of monolayer CrTe$_2$ in the FM phase. Blue (red) lines correspond to majority (minority) spin states. The inset in (a) shows the DFT Fermi surface. Points show parts of the Fermi surfaces connected by nesting vectors ${\mathbf q}_M$ and ${\mathbf q}_K$; the wave vectors ${\mathbf Q}_M$ and ${\mathbf Q}_K$ are shown for comparison.} 
\label{Fig:bands_1_layer_FM}
\end{figure}

In Fig. \ref{Fig:bands_1_layer_FM} we show the band structure (cf. Refs. \cite{li_tunable_2021,zhang_room-temperature_2021}) and Fermi surfaces for the FM phase of the monolayer CrTe$_2$. The nesting vectors ${\mathbf q}_M$ and ${\mathbf q}_K$ along the $\Gamma-M$ and $\Gamma-K$ directions are close to the vectors ${\mathbf Q}_M=\pi(1,1\sqrt{3})/a$ and ${\mathbf Q}_K$. By parameterizing the dispersion near the respective pairs of points of the Fermi surface by $e_{\mathbf k}=v_{F\sigma}\tilde{k}_x+\tilde{k}_y^2/(2m_\sigma)$ where $\tilde{k}_{x,y}$ are local coordinates, assuming opposite Fermi velosities and masses of different spin states, we find at small temperatures the contribution to the exchange interactions
\begin{align}
J_{\mathbf{q}}\propto \frac{\Delta^2}{|v_{F\uparrow}|+|v_{F\downarrow}|} \ln\frac{(|v_{F\uparrow}|+|v_{F\downarrow}|)^2}{|v_{F\uparrow}v_{F\downarrow}(\rho_{\uparrow}-\rho_{\downarrow})|}
\end{align}
where $\rho_\sigma=1/ (m_\sigma v_{F\sigma})$ are the curvatures of the Fermi surface sheets, $\Delta$ is the sverage spin splitting. Although Fermi velocities are somewhat different at the respective sheets of Fermi surfaces, this does not prevent the nesting property, as long as the difference of curvatures $\rho_{\uparrow}-\rho_{\downarrow}$ remains small close to the considered points, connected by the vectors ${\mathbf q}_M$ and ${\mathbf q}_K$.

\section{Energy mapping of monolayer system}

\begin{figure}[t]
\centering
\includegraphics[width=1.0\linewidth]{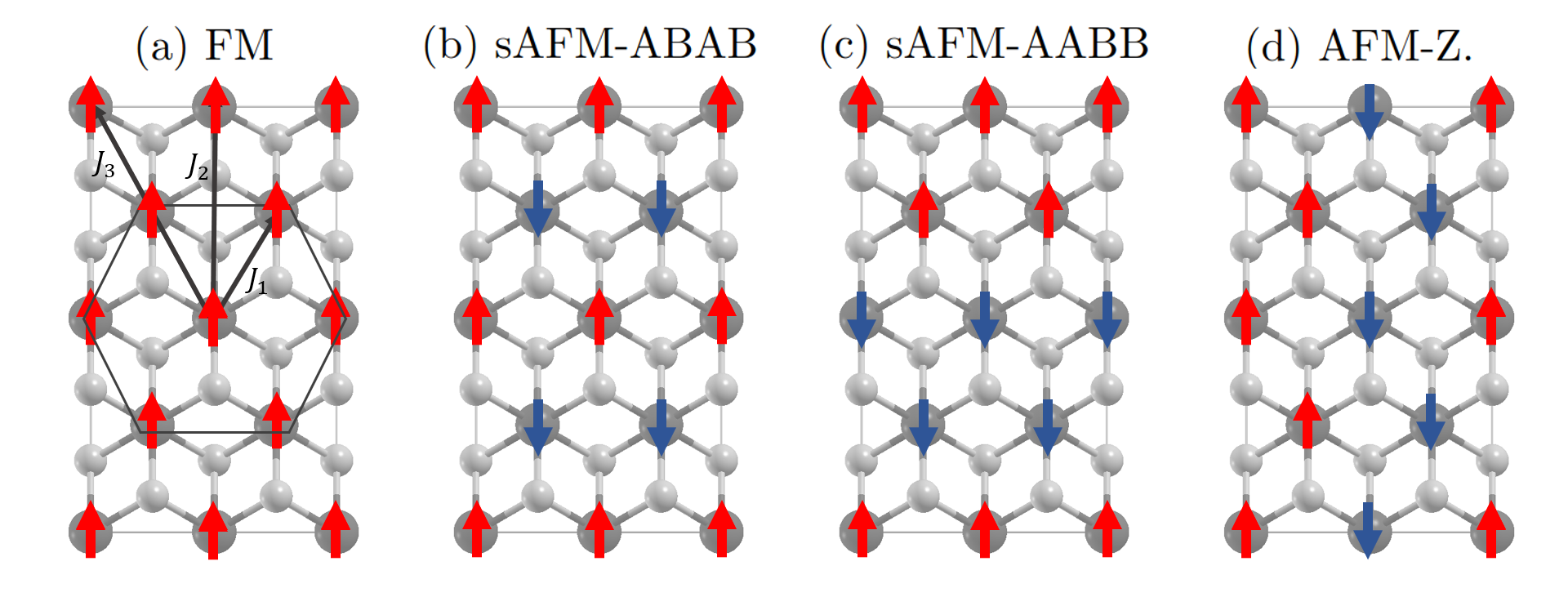}
\caption{(Color online) Schematic spin configuration for (a) FM
(b) AFM-ABAB 
(c) AFM-AABB 
(d) AFM-zz. 
}
\label{Fig:emap}
\end{figure}  
 
To estimate exchange interactions from the energies of various magnetic states, we performed calculations on the $2 \times 2 \sqrt{3}$ supercell, as shown in Fig. \ref{Fig:emap}. 
We consider 
4 different spin configurations to estimate the exchange parameters. For all configurations the same lattice structure was used. 
Each atom has six nearest neighbors, six next nearest neighbours and six next to next nearest neighbours and we have eight atoms in the cell. To obtain corresponding energies it is enough to consider one of the atoms and multiply the result by 8 atoms.

The obtained DFT energies are 
\begin{align}
E_{\mathrm{FM}} =E_0-24 (J_1+J_2+J_3)S^2&=128.968~{\mathrm eV}\notag\\
E_{\mathrm{AFM-ABAB}}  =E_0+8(J_1+J_2 -3 J_3)S^2&=129.003~{\mathrm eV}\notag\\
E_{\mathrm{AFM-AABB} }  =E_0-8(J_1-J_2- J_3)S^2&=130.197~{\mathrm eV}\notag\\
E_{\mathrm{AFM-zz} }  =E_0+8(\comAK{J_1}-J_2+J_3)S^2&=129.051~{\mathrm eV}\notag
\end{align}
from which we get exchange parameters, presented in Table I.


\end{document}


\preprint{APS/123-QED}
\title{Erratum: Magnetic properties of monolayer, multilayer, and bulk CrTe$_2$}
\author{A. A. Katanin}
\affiliation{Center for Photonics and 2D Materials, Moscow Institute of Physics and Technology, Institutsky lane 9, Dolgoprudny, 141700, Moscow region, Russia}
\affiliation{M. N. Mikheev Institute of Metal Physics of the Ural Branch of the Russian Academy of Sciences, S. Kovalevskaya Street 18, 620990 Yekaterinburg, Russia}
\author{E. M. Agapov}
\affiliation{Ludwig-Maximilians-Universität München, Geschwister-Scholl-Platz 1, 80539, Munich, Germany}
\affiliation{Technical University of Munich, School of Natural Sciences, Arcisstraße 21, 80333, Munich, Germany}
\date{\today}
\begin{abstract}
In our study of Ref. \cite{OurPaper} we found for the monolayer CrTe$_2$ the most preferable 120$\degree$ spin spiral structure. While this order is expected for freely suspended CrTe$_2$, for experimentally realized single layer CrTe$_2$ on a substrate, which is characterized by larger lattice constant, the momentum dependence of the susceptibilities drastically changes. In particular, the ferromagnetic ground state is expected for this compound, in agreement with the experimental data. 
\end{abstract}
\maketitle

{

Our study of monolayer CrTe$_2$ in Ref. \cite{OurPaper} used the lattice parameter $a=3.71$~\AA, which is obtained from the relaxation of crystal structure in the ferromagnetic phase, and coincides with the lattice parameter obtained by similar procedure in Ref. \cite{abuawwad_noncollinear_2022}. The study of Ref. \cite{OurPaper} predicted the 120$\degree$ order in this monolayer compound, seemingly contradicting the experimental data  \cite{zhang_room-temperature_2021, meng_anomalous_2021, liu_wafer-scale_2023}, where the ferromagnetic state of monolayer CrTe$_2$ was reported. 

In this Erratum we emphasize that the origin of this difference can be attributed to the difference of lattice parameters owing to the substrate, similarly to the previous DFT studies \cite{wu_-plane_2022,zhu_insight_2023,gao_thickness_2021,lv_strain-controlled_2015,AFM2}, which considered the transition frm AFM-zz to FM state with increase of the lattice parameter. Indeed, the experimentally measured lattice parameter of CrTe$_2$ on graphene \cite{zhang_room-temperature_2021,CrTe2exp} is $a=3.81$~\AA, while Al$_2$O$_3$ substrate yields even larger lattice parameter $a=3.84$~\AA~\cite{liu_wafer-scale_2023}. Although these parameters are only $3\%$-$4\%$ larger than that obtained from the relaxation, they lead to quite different results in the DFT+DMFT approach.

\begin{figure}[b]
\centering
\includegraphics[width=0.62\linewidth,angle=-90]{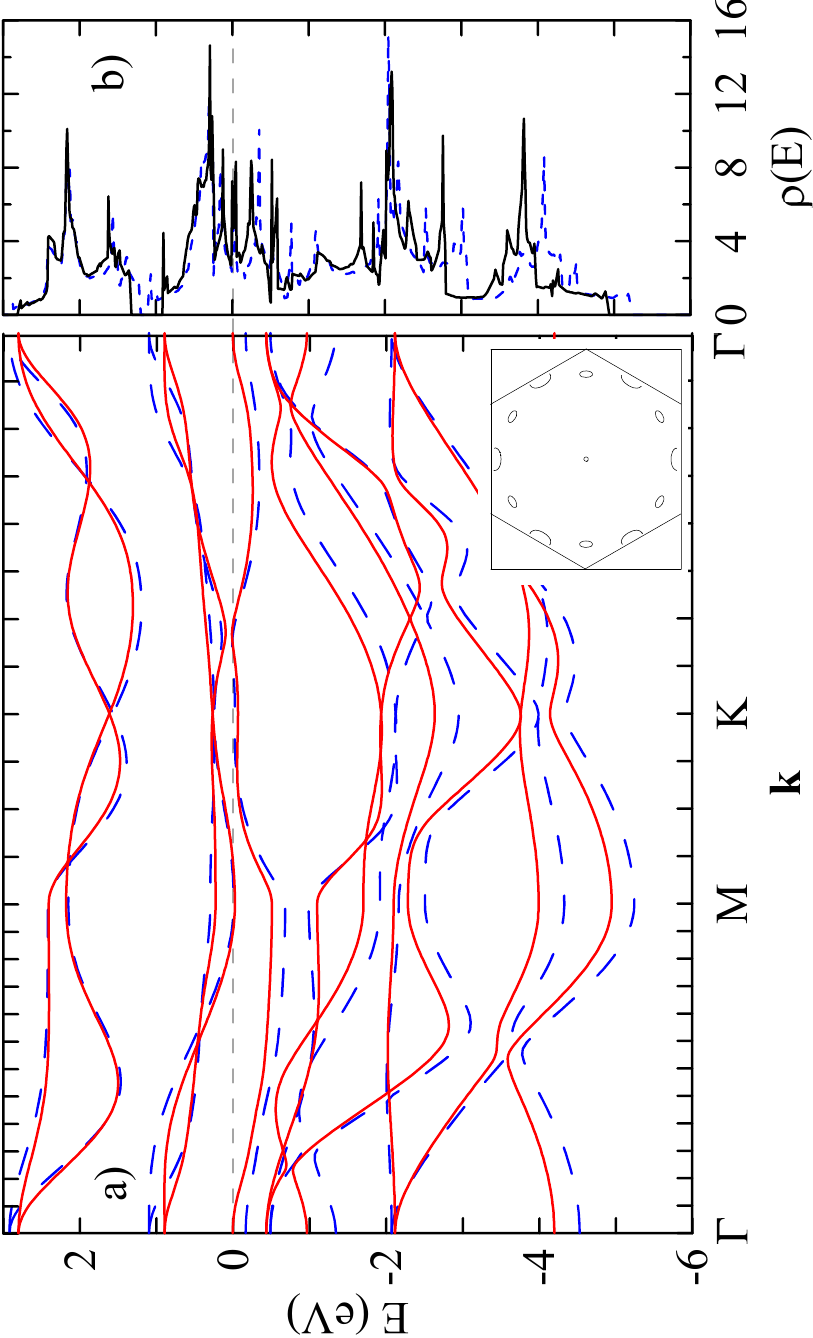}
\caption{(Color online) Band structure (a) and DFT density of states (b) of the monolayer CrTe$_2$ in paramagnetic phase for $a=3.81$\AA~(solid lines), compared to that for $a=3.71$\AA~ (dashed lines), presented in Ref. \cite{OurPaper}. The inset in (a) shows the DFT Fermi surface for $a=3.81$\AA.}
\label{Fig:bands_1_layer}
\end{figure}

The DFT band structure of the monolayer CrTe$_2$ with the lattice parameter $a=3.81$~\AA~ in the paramagnetic state is shown and compared to that of our study \cite{OurPaper} in Fig. \ref{Fig:bands_1_layer}. One can see that the band structure is only slightly changed, but one of the bands crosses Fermi level near $\Gamma$ point. This yields additional electron pocket at the center of the Brillouin zone (see the inset of Fig. \ref{Fig:bands_1_layer}). As a result of this, the peak of the density of states at the Fermi level becomes stronger with increase of the lattice constant. Apart from that, the geometry of the DFT Fermi surface is changed, in particular the nesting at the wave vector ${\mathbf Q}={\mathbf Q}_K$, which was present for  $a=3.71$~\AA, is deteriorated. 

\begin{figure}[b]
\centering
\includegraphics[width=0.47\linewidth,angle=-90]{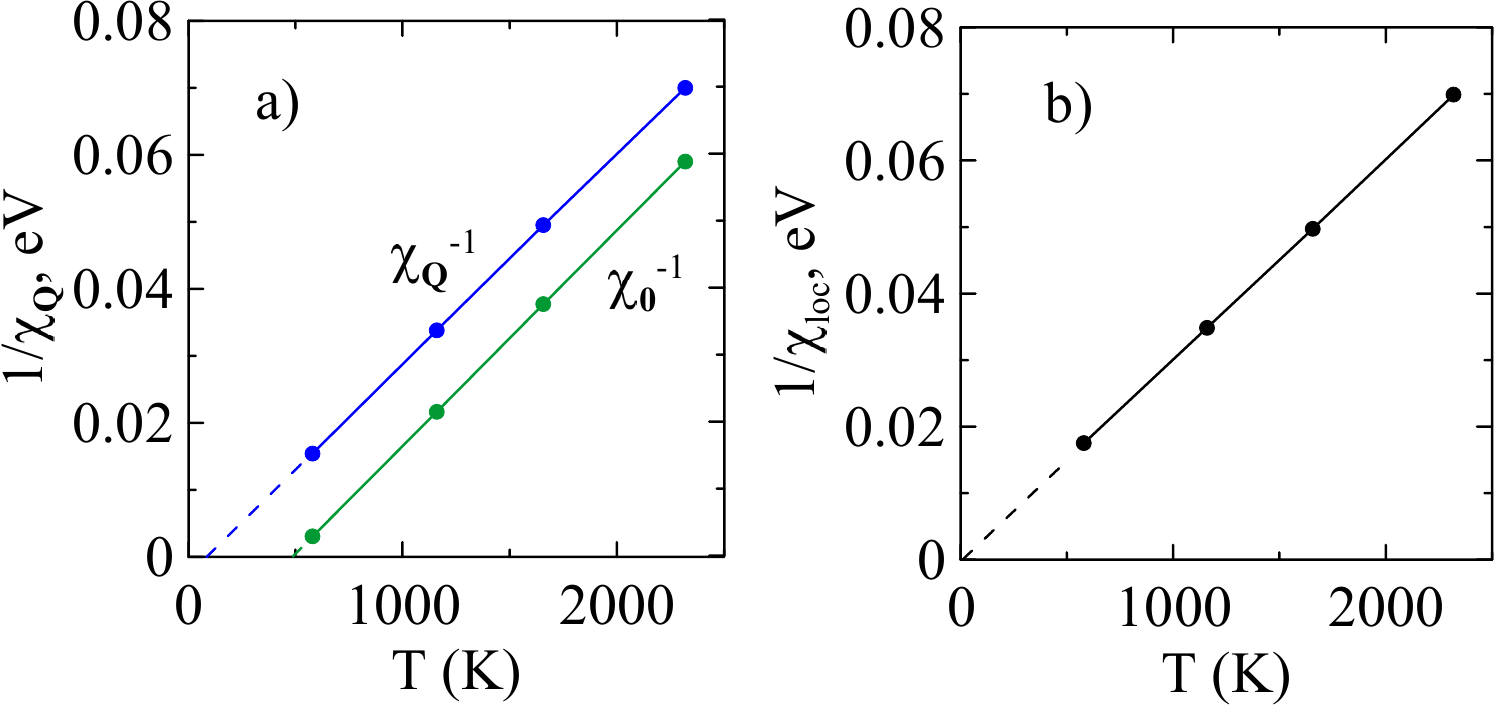}
\caption{(Color online) Inverse uniform and staggered, ${\mathbf Q}={\mathbf Q}_K$ (a) and local (b) spin susceptibility of the monolayer CrTe$_2$ with $a=3.81$~\AA~ in DFT+DMFT approach. Dashed lines show extrapolation to the low-temperature region. }
\label{Fig:chi_1_layer}
\end{figure}

The temperature dependence of the inverse uniform and staggered magnetic susceptibilities for $a=3.81$~\AA~is shown in Fig. \ref{Fig:chi_1_layer}. Despite the slope of these temperature dependencies (i.e. the respective magnetic moments) not changing much in comparison to $a=3.71$\AA, the inverse susceptibilities $\chi_0^{-1}$ and $\chi_{{\mathbf Q}_K}^{-1}$ change their order: we find $\chi_{{\mathbf q}=0}>\chi_{{\mathbf Q}_K}$ and the respective Curie temperature $T_C^{\rm DMFT}\simeq 500$~K in the DMFT approach.

\begin{figure}[t]
\centering
\includegraphics[width=0.9\linewidth]{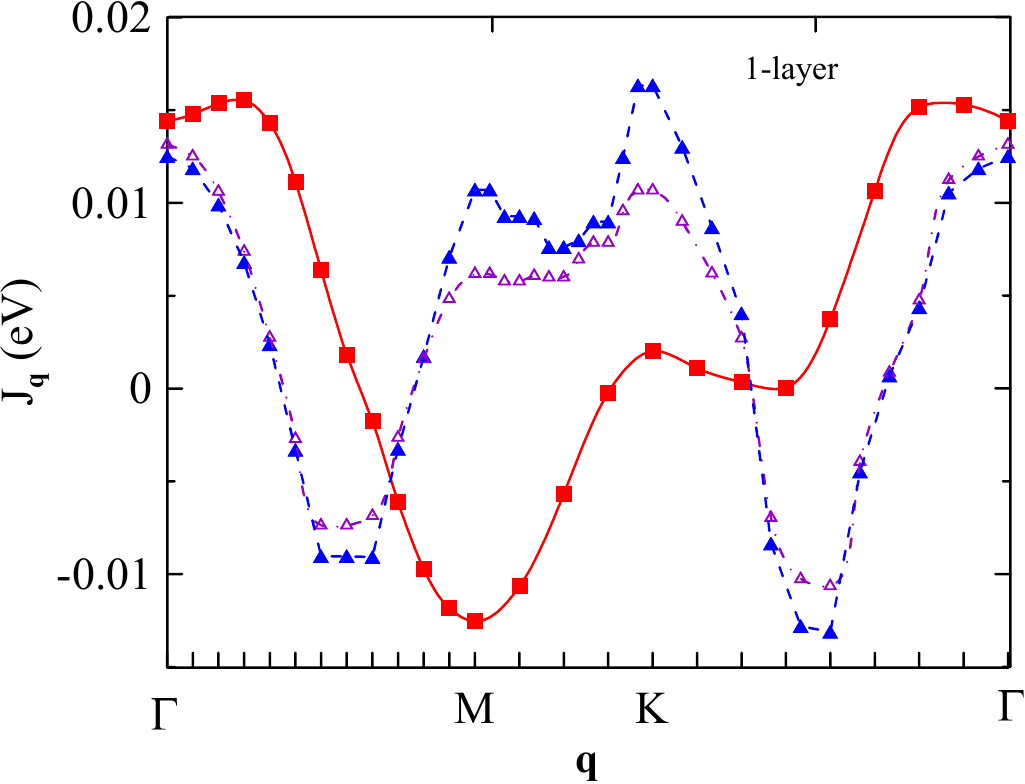}
\caption{(Color online) Exchange interactions in monolayer CrTe$_2$ with $a=3.81$~\AA~at $\beta=20$~eV$^{-1}$ ($T=580$~K). Red solid line (squares) corresponds to DFT+DMFT approach in paramagnetic phase, blue dashed line (solid triangles) is the result of the DFT approach in FM phase. 
For comparison, the result of DFT approach at $\beta=10$~eV$^{-1}$ \comAK{($T=1160$~K)} is shown by dot-dashed violet line with open triangles
}
\label{Jq_1_layer}
\end{figure}

Magnetic exchange interactions of the monolayer CrTe$_2$ at $\beta=20$~eV$^{-1}$ are shown in Fig. \ref{Jq_1_layer} (they also rather weakly depend on temperature). The maximum of exchange interaction in DFT+DMFT approach is obtained close to the $\Gamma$ point (in contrast to the K point at $a=3.71$~\AA), which corresponds to the incommensurate state with small wave vector.  
Similarly to the results of Ref. \cite{OurPaper}, we find that the peak of exchange interaction $J_{\mathbf q}$ at $\mathbf{q}=\mathbf{Q}_K$ in DFT approach sharply increases with a decrease of temperature, due to nesting of the Fermi surface pockets with the wave vector close to $\mathbf{Q}_{K}$ in the magnetically ordered state. Therefore, in contrast to the DFT approach, the DFT+DMFT approach yields magnetic instability with a small wave vector at low temperatures.

We note that previously the ferromagnetic state was claimed to be stable in the DFT approach at a sufficiently large lattice parameter $a$, on the basis of the energy comparison to the AFM-zz state. However, from the present analysis it is tempting to conclude, that the ferromagnetic phase is unstable in DFT towards incommensurate order with ${\mathbf Q}={\mathbf Q}_K$ for $a=3.81$\AA. We have verified that this is likely related to the change of Te positions when performing relaxation in the ferromagnetic phase. Indeed, comparing for $a=3.81$\AA~the energies of various magnetic configurations shown in Fig. 15 of the paper \cite{OurPaper} for Te positions obtained in the ferromagnetic phase, we find  $E_{\rm ABAB}<E_{\rm FM}<E_{\rm AFM-zz}$,  while for Te positions obtained in the paramagnetic phase, we find  $E_{\rm FM}<E_{\rm AABB}<E_{\rm ABAB,AFM-zz}$,
which we believe more accurately reproduces comparison of the energies of various magnetic configurations. The obtained exchange interactions with tellurium  positions obtained within paramagnetic phase are shown in Table I. One can see, that the difference between the DFT magnetic force theorem (MFT) and DFT Energies approaches is mainly in the sign of the nearest neighbor interaction, which appears to be sensitive to Te positions.

\vspace{0.3cm}

\begin{table}[h!]
\begin{tabular}{lrrr}
Method                     & $\ \ J_1$\ \  & $J_2$\ \   & $J_3$ \ \ \\
\hline
DFT MFT & $-0.87$    & $2.67$  & $1.63$ \\
DFT Energies  & $3.56$   & $2.84$ &  $1.42$    \\
DFT+DMFT                & \ \ $2.77$    & $1.00$  & $-1.03$ \\
\end{tabular}
\caption{Exchange interactions (in meV) in monolayer CrTe$_2$ for $a=3.81$\AA~ between nearest ($J_1$), next nearest ($J_2$) and next to next nearest ($J_3$) neighbours within the magnetic force theorem (MFT) approach, comparison of energies of various phases in DFT and DFT+DMFT approach in paramagnetic phase. DFT MFT and DMFT exchange interactions are calculated at $\beta=20$~eV$^{-1}$ \comAK{($T=580$~K)}.} 
\label{Table:exchenge_monolayer}
\end{table}

{\it Acknowledgements}. DMFT calculations were supported by the Russian Science Foundation (project 24-12-00186). The DFT calculations in the paramagnetic phase were carried out within the framework of the state assignment of the Ministry of Science and Higher Education of the Russian Federation for the IMP UB RAS.